\pdfoutput=1
\documentclass[12pt,preprint]{aastex}
\usepackage{amsmath}
\begin{document}
\newcommand{\up}[1]{\ifmmode^{\rm #1}\else$^{\rm #1}$\fi}
\newcommand{\zdot}{\makebox[0pt][l]{.}}
\newcommand{\upd}{\up{d}}
\newcommand{\uph}{\up{h}}
\newcommand{\upm}{\up{m}}
\newcommand{\ups}{\up{s}}
\newcommand{\arcd}{\ifmmode^{\circ}\else$^{\circ}$\fi}
\newcommand{\arcm}{\ifmmode{'}\else$'$\fi}
\newcommand{\arcs}{\ifmmode{''}\else$''$\fi}

\title{The Araucaria Project. Multi-band calibrations of the TRGB absolute magnitude}

\author{Marek G{\'o}rski}
\affil{Universidad de Concepci{\'o}n, Departamento de Astronomia,
Casilla 160-C, Concepci{\'o}n, Chile}
\affil{Millennium Astrophysical Institute, Santiago, Chile}
\authoremail{mgorski@astro-udec.cl}

\author{Grzegorz Pietrzy{\'n}ski}
\affil{Nicolaus Copernicus Astronomical Center, Polish Academy of Sciences, Bartycka 18, 00-716, Warsaw, Poland}
\affil{Universidad de Concepci{\'o}n, Departamento de Astronomia,
Casilla 160-C, Concepci{\'o}n, Chile}
%\authoremail{pietrzyn@camk.edu.pl}

\author{Wolfgang Gieren}
\affil{Universidad de Concepci{\'o}n, Departamento de Astronomia,
Casilla 160-C, Concepci{\'o}n, Chile}
\affil{Millennium Astrophysical Institute, Santiago, Chile}
%\authoremail{wgieren@astro-udec.cl}

\author{Dariusz Graczyk}   
\affil{Nicolaus Copernicus Astronomical Center, Polish Academy of Sciences, Bartycka 18, 00-716, Warsaw, Poland}
\affil{Millennium Astrophysical Institute, Santiago, Chile}
\affil{Universidad de Concepci{\'o}n, Departamento de Astronomia,
Casilla 160-C, Concepci{\'o}n, Chile}
%\authoremail{darek@astro-udec.cl}

\author{Ksenia Suchomska}  
\affil{Warsaw University Observatory, Al. Ujazdowskie 4, 00-478, Warsaw,
Poland}
\affil{Universidad de Concepci{\'o}n, Departamento de Astronomia,
Casilla 160-C, Concepci{\'o}n, Chile}
%\authoremail{ksenia@astrouw.edu.pl}

\author{Paulina Karczmarek}
\affil{Warsaw University Observatory, Al. Ujazdowskie 4, 00-478, Warsaw,  
Poland}
%\authoremail{pkarczma@astrouw.edu.pl}

\author{Roger E. Cohen}
\affil{Space Telescope Science Institute, 3700 San Martin Drive, Baltimore, MD 21218, USA}
\affil{Universidad de Concepci{\'o}n, Departamento de Astronomia, Casilla 160-C, Concepci{\'o}n, Chile}

\author{Bart\l{}omiej Zgirski}
\affil{Nicolaus Copernicus Astronomical Center, Polish Academy of Sciences, Bartycka 18, 00-716, Warsaw, Poland}
%\authoremail{bzgirski@camk.edu.pl}

\author{Piotr Wielg{\'o}rski}
\affil{Nicolaus Copernicus Astronomical Center, Polish Academy of Sciences, Bartycka 18, 00-716, Warsaw, Poland}
%\authoremail{pwielgorski@astrouw.edu.pl}

\author{Bogumi\l{} Pilecki}   
\affil{Nicolaus Copernicus Astronomical Center, Polish Academy of Sciences, Bartycka 18, 00-716, Warsaw, Poland}
\affil{Universidad de Concepci{\'o}n, Departamento de Astronomia,
Casilla 160-C,Concepci{\'o}n, Chile}

\author{M{\'o}nica Taormina}   
\affil{Nicolaus Copernicus Astronomical Center, Polish Academy of Sciences, Bartycka 18, 00-716, Warsaw, Poland}

\author{Zbigniew Ko\l{}aczkowski}
\affil{Nicolaus Copernicus Astronomical Center, Polish Academy of Sciences, Bartycka 18, 00-716, Warsaw, Poland}
%\authoremail{}

\author{Weronika Narloch }
\affil{Universidad de Concepci{\'o}n, Departamento de Astronomia, Casilla 160-C, Concepci{\'o}n, Chile}
\affil{Millennium Astrophysical Institute, Santiago, Chile}
\affil{Nicolaus Copernicus Astronomical Center, Polish Academy of Sciences, Bartycka 18, 00-716, Warsaw, Poland}
%\authoremail{}

\begin{abstract}
We present new empirical calibrations of the absolute magnitude of the tip of the red giant branch (TRGB) in the optical I and near-infrared J, H, and K bands in terms of the 
(V-K)$_0$, (V-H)$_0$, and (J-K)$_0$ colors of the red giant branch. Our calibrations are based on the measurements in 19 fields in the Large and Small Magellanic Clouds, which span a wide (V-K)$_0$ color range of the brightest part of the red giant branch. We use a simple edge detection technique based on the 
comparison of the star count difference in two adjacent bins with the estimated Poisson noise. Further, we include the reddening and geometrical corrections, as well as the precise and accurate to 2\% distance to the Large Magellanic Cloud. The calibration based on a (V-K) colors can be a robust tool to calculate with a great precision the absolute magnitude of the TRGB. 
\end{abstract}


\section{Introduction}
In recent years a lot of attention was put on distance measurements, and on improving the calibration
of the distance scale (Riess et al. 2016; Beaton et al. 2016). In our long-term Araucaria Project we have investigated and applied most of the precise distance measurement methods, like the mean brightness of the red clump stars 
(Pietrzy{\'n}ski \& Gieren 2002, Pietrzy{\'n}ski et al. 2010), Cepheid period-luminosity (P-L) relation 
(Gieren et al. 2005; Zgirski et al. 2017; Wielg{\'o}rski et al. 2017), RR Lyrae stars (Szewczyk et al. 2009;
Karczmarek et al. 2017), late-type 
eclipsing binaries (Pietrzy{\'n}ski et al. 2009a; Graczyk et al. 2018) and blue supergiants (Urbaneja et al. 2008; Berger et al. 2018). 
In this paper we continue our investigations on the tip of the red giant branch (TRGB) method (Pietrzy{\'n}ski et al. 2009b; G{\'o}rski et al. 2011; G{\'o}rski et al. 2016).

The TRGB is the sharp cut-off on the color-magnitude diagram occurring at the bright end of the 
red giant branch (RGB). It marks the final stage of the evolution of stars during the RGB phase, which is 
terminated by a helium flash. Because all low mass stars have a similar I band brightness just before the helium flash, 
and this brightness very weakly depends on the stellar mass and age for old metal-poor stellar populations, the I band magnitude of the TRGB can be used as 
a standard candle (Madore \& Freedman 1995; Barker et al. 2004).

The idea of using the TRGB to measure the distance was first used by Baade (1944a). With red-sensitive 
photographic plates he observed the central region of the Andromeda galaxy and its two companion galaxies, M32 
and NGC 205. He noticed that the brightest red giants in all three galaxies have the same magnitude and color. 
Moreover, he stated that the galaxies NGC 147 and NGC 185 should be at the same distance as the Andromeda system, 
since the brightest red stars in those galaxies also have a similar magnitude as the stars in the Andromeda system 
(Baade 1944b). In 1971 Sandage found that the brightest red stars in the IC 1613 galaxy have the same absolute magnitude
 as in the M31 and M33 galaxies. 

During the next decades more sophisticated techniques to measure the brightness of the TRGB were developed, and with 
the arrival of CCD measurements the TRGB brightness was used to determine the distances to almost all Local 
Group galaxies. With those measurements many different investigations were conducted to check the reliability 
of the TRGB method. It became clear that the I band TRGB brightness depends on the red giant branch metallicity 
at the level of 0.1 mag. This dependence was calibrated by Da Costa \& Armandroff in 1990. In the same paper, 
the authors presented a relation to calculate the metallicity from the (V-I) color of the RGB, which was the first 
indirect calibration of the TRGB absolute I band brightness from the color of the RGB. 

In 1993, Lee, Freedman and Madore compared the distances obtained with the TRGB I band brightness with distances 
obtained with the Cepheid P-L relation and RR Lyrae stars for 10 Local Group galaxies. The obtained distances agree 
within 0.2 mag, proving that the TRGB method can be successfully applied to measure the distance. Until now, 
the TRGB method was applied to determine the distances to more than 300 galaxies up to 16 Mpc (Jacobs et al. 2009; 
Tully et al. 2016; Hatt et al. 2018).

Since metallicity measurements are scarce, most calibrations are based on the (V-I) color of the red giant branch 
(Rizzi et al. 2007; Jacobs et al. 2009). Very recently Jang \& Lee (2017) calibrated the optical I band absolute 
magnitude of the TRGB in terms of the F814W - F555W color, which is the ACS/WFC Hubble Space Telescope equivalent 
of the (V-I) color. The zero point of this calibration is based on two distance anchors, NGC 4258 (M106) 
and the LMC. The advantage of this approach lies in the fact that the precise geometrical distances to 
both galaxies are known (Herrnstein et al. 1999; Pietrzy{\'n}ski et al. 2013)

One of the biggest contributions to the total uncertainty of the distances measured with the TRGB in the I band, comes from 
the interstellar extinction. Usually only the Galactic foreground reddening is taken into account and 
the internal extinction is assumed to be negligible. This approach is justified as long, as the observations are 
performed in the halo of the galaxies, which is presumably dust free. 
In some cases, this assumption can lead to a systematic errors. In recent years 
our group measured the distance and reddening to a number of Local Group galaxies based on the
near-infrared photometry of Cepheid variables. We found that the total reddening tends to by systematically
higher compared to the Galactic foreground reddening. If the TRGB stars are affected by the internal 
reddening in a similar way as Cepheids, the reddening underestimation 
by 0.05 mag will lead to a distance moduli overestimation in the I band by 0.06 mag. 
(Gieren et al. 2005; Soszy{\'n}ski et al. 2006; Pietrzy{\'n}ski et al. 2006; 
Gieren et al. 2006; Gieren et al. 2008; Gieren et al. 2009; 
Gieren et al. 2013;  Zgirski et al. 2017)

The effect of reddening can be reduced by utilising near-infrared bands. In 2004 Valenti, Ferraro \& Origlia 
calibrated the TRGB infrared J, H, and K absolute brightnesses in terms of metallicity. 
In the last 10 years, this calibration was applied to measure the distance only to a few galaxies 
(G{\'o}rski, Pietrzy{\'n}ski, Gieren 2011). The main disadvantage of this method is a strong sensitivity of the 
near-infrared bands to metallicity (a 0.1 dex metallicity difference changes the brightness of the TRGB in the K band 
by 0.058 mag), in tandem with scarce spectroscopic metallicity measurements (Cohen et al. 2017).

In our last paper (G{\'o}rski et al. 2017) we empirically confirmed that the metallicity dependent calibration 
of Valenti et al. (2004) leads 
to systematic errors at the level of 0.2 mag, if applied to the Large and Small Magellanic Clouds (LMC and SMC, respectively).
This problem was already confirmed by many theoretical studies (Barker et al. 2004; Salaris \& Girardi 2005; 
Serenelli et al. 2017) and is caused by population effects, namely the age and chemical composition of the red giants. 
In contrast to metallicity dependent calibrations, the color - TRGB absolute magnitude relations are much less affected 
by this systematic error (Serenelli et al. 2017; G{\'o}rski et al. 2017). 

Recently, both theoretical and empirical calibrations of the infrared TRGB brightness in terms of the (J-K) and (J-H) 
colors were published. 
Serenelli et al. (2017) provided a solid theoretical calibration based on a careful stellar modeling. 
Hoyt et al. (2018) and Madore et al. (2018) derived the empirical near-infrared calibrations of the TRGB based on the LMC distance and slope of the TRGB observed 
in the IC 1613 dwarf galaxy. 

Motivated by these results, we decided to independently calibrate the TRGB absolute magnitude, taking advantage of 
the wide color spread observed in different fields in the LMC and SMC, and the accurate geometrical distances to both galaxies
measured recently with the eclipsing binaries method. 
While our focus is on the (V-K) color which allows one
to calculate the absolute brightness of the TRGB with a great precision, 
we also investigate the (V-H) and (J-K) color calibrations. 

During the last decade the TRGB method and Cepheid P-L relation were used to calibrate the absolute 
magnitudes of supernovae Ia (SNe Ia) in nearby galaxies, and determine the value of the Hubble constant. In comparison with the Cepheid P-L relation, the TRGB was used to measure distance to a much larger 
number of galaxies, including galaxies that lack young standard candles like classical Cepheids. Cepheid distance measurements 
are also affected by numerous systematic errors including reddening, a possible metallicity dependence, or a
nonlinearity of the P-L relation (Kodric et al. 2015).
Therefore the TRGB method is an important tool that can complement and perhaps improve the determination of the 
Hubble parameter. 

Our paper is organised as follows. The data sources, edge detection technique, and the TRGB and color measurements are described 
in the following section. In section 3 we present the resulting calibrations. Results are discussed in section 4. 
Finally we present a summary and conclusions. 

\section{Data analysis}
The optical V and I band photometry of the stars in the Large and Small Magellanic Clouds was acquired from the photometric maps 
of the OGLE-III survey (Udalski et al. 2008a, 2008b). The OGLE-III photometric maps were obtained with the 1.3 m Warsaw telescope 
located at the Las Campanas Observatory. The telescope was equipped with a mosaic camera with a 0.26 arcsec pixel scale. The V 
and I band magnitudes were calibrated onto the standard system using Landolt standards. The source of the near-infrared J, H, and K band brightnesses is the 
IRSF Magellanic Clouds Point Source Catalog (Kato et al. 2007).  
The IRSF is a 1.4 m telescope, located at the South African Astronomical Observatory (SAAO), equipped with the SIRIUS camera 
(0.45 arcsec pixel scale). 
The photometric system consists of three near-infrared filters similar to the 2MASS and UKIRT photometric systems. This allowed 
us to transform the magnitudes onto the 2MASS NIR photometric system following the procedure described by Kato et al. (2007). 
The optical V and I band catalog of OGLE-III was crossmatched with the IRSF near-infrared catalog based 
on the provided coordinates. The statistical photometric uncertainty of stars that were used in our analysis is 0.03 mag for the 
V band, and below 0.02 mag for the I, J, H, and K bands. 
To estimate the systematic error on the photometry, we decided to crossmatch the brightest stars with the 2MASS catalog 
(Skrutskie et al. 2006). The mean magnitude difference between our transformed IRSF brightnesses and the 2MASS catalog is below 
0.01 mag for all bands. 

The photometric data were divided into 25 fields covering the central regions of the LMC and 8 fields in the central part 
of the SMC. The size of each field, both in the LMC and SMC is  35 arcmin $\times$ 35 arcmin. From the total of 33 fields, 
only 14 and 5 fields in the LMC and SMC respectively were used in the final analysis. The reason for this selection 
is  described later in this section, and discussed in the final part of the paper. The coordinates of the analysed fields 
and the names are given in Table 1. The names of the fields are consistent with the OGLE-III catalog naming convention. 

For each field, the color-magnitude diagram was created, and red giant branch stars were selected based on 
the (V-K) color, and corresponding K band brightness. 
Figure 1 presents an example of the red giant branch stars selected on the color-magnitude diagram for field LMC127.   

\subsection{Edge detection techniques}

In order to secure a precise and accurate measurement of the TRGB brightness in each field, we decided to use different techniques 
and investigate the results of the edge detection methods. 
The first method we used is the the Sobel filter, described by Lee, Freedman \& Madore (1993) , and later improved 
by Sakai, Madore \& Freedman (1996).
The Sobel filter is operating on the Gaussian-smoothed luminosity function $\Phi(m)$, which follows the expression: 

\begin{equation}
\Phi(m)=\sum_i^N \frac{1}{\sqrt{2\pi}\sigma_i}\exp\left[-\frac{(m_i-m)^2}{2\sigma^2_i}\right] \mathrm ,
\label{eq:sobel}
\end{equation}
where m$_i$ is the magnitude of the $i$-th star, $\sigma_i$ is the $i$-th star photometric error, and $N$ is the total number 
of stars in the sample. 
The Sobel filter answer $E(m)$ is defined as:
\begin{equation}
E(m) = \Phi (m+a) - \Phi (m-a) \mathrm ,
\end{equation}
where $a$ is the mean photometric error for all the stars within magnitudes $m - 0.05$  and $m + 0.05$  mag. The brightness 
corresponding to the highest value of the Sobel filter answer is the  brightness of the TRGB. 
Given it's simplicity the Sobel filtering technique has been widely adopted, and was employed by us in our previous papers.

While the Sobel filter is sufficient for most of the applications, in the case of some fields in the LMC and SMC 
it is difficult to measure the TRGB brightness, because in proximity of the expected cut-off on the luminosity 
function, the Sobel filter answer shows multiple peaks.

The second implemented method of the TRGB brightness measurement is the Maximum Likelihood Algorithm (MLA), introduced 
by Mendez et al. (2002) and later improved by Makarov et al. (2006). 
In contrast to the previously described method, in the MLA  a theoretical predefined luminosity function is fitted to the 
observed distribution of the stars. Additionally, this method incorporates photometric errors and a completeness function. 
In this approach, the luminosity function is assumed to be a simple power-law with a cut-off for the TRGB brightness: 

\begin{equation}
 \psi = \begin{cases}
    10^{a(m-m_\mathrm{TRGB})+b} & \mbox{for } m-m_\mathrm{TRGB} \geq 0\\
    10^{c(m-m_\mathrm{TRGB})} & \mbox{for } m-m_\mathrm{TRGB} < 0
  \end{cases}
\end{equation}

Calculating the Maximum Likelihood allows to estimate the TRGB brightness ($m_{\mathrm{TRGB}}$) and luminosity function slope parameters 
($a$,$b$ and $c$ of Eq. 3). 
This method proved to be especially convenient if the part of the luminosity function in the proximity of the TRGB is poorly 
populated, or it approaches the photometric limit.
Unfortunately, for many analysed fields in the LMC and SMC the calculated TRGB brightness differs by more than 1 mag from the 
expected value. The cause of this
 discrepancy is connected with an oversimplified model of the luminosity function in the case of the LMC and SMC. 

Our final approach is based on the Poisson noise weighted star counts difference in two adjacent bins 
(hereinafter the PN method). The response of this filter is defined for any magnitude ($m$) with desired resolution by the 
following equation:

\begin{equation}
PN(m)=\frac{(N_U-N_L)}{\sqrt{(N_U+N_L)/2}} \mathrm .
\end{equation}
$N_U$ is the number of stars in the bin of magnitude from $m$ to $m+\mu$, and $N_L$ is the number of stars 
in the bin of magnitude from $m-\mu$ to $m$. The $\mu$ parameter value
 in our implementation was set between 0.1 and 0.3 mag, and the resolution of the calculations was set to 0.01 mag.
This method was introduced with a slightly different formula by Madore et al. 2009 to 
statistically estimate the significance of the Sobel filer [-1, 0, +1] kernel response.
In our application we convolved the PN filter answer with a Gaussian function:
 
 \begin{equation}
g(m)= \exp\left[-\frac{(m-m_0)^2}{2\sigma^2}\right] \mathrm.
\end{equation}
 
This procedure is used to smooth the response of the filter, which slightly improves the accuracy of the measurements, as long as the $\sigma$ value does not significantly exceeds mean photometric error of the stars. To obtain the desired smoothing we used $\sigma$=0.03 mag. 

The main advantage of this method is the clarity of the response. Compared with the Sobel filter, the main peak is usually 
very distinctive, and the value of the response has a clear interpretation,  since it corresponds to expected variations 
of the star counts number in the selected range of magnitudes. In the following subsection we present some important 
properties of this filter. 

\subsection{PN filter properties}
Figure 2 presents examples of the PN filter response on the artificially created power law distribution of star magnitudes. 
The power law is described by Eq. 3, with the edge corresponding to the TRGB at 10 mag. Parameters of the distribution 
are $a$=0.30, $b$=0.30 and $c$=0.35, and correspond to the typical values in the LMC. Those values were calculated with the 
MLA technique. 
It is clear that for the power law distribution, with increasing bin size $\mu$ the PN response value has intrinsic rising trend 
with increasing magnitude, which can lead to a systematic measurement error. This was the main reason for us to limit the value of the $\mu$ parameter 
to 0.3 mag. To investigate for any possible systematic errors connected with the properties of the PN filter, we performed 
a series of simulations. Artificial luminosity function described by Eq. 3 was created with parameters 
$m_\mathrm{TRGB} = 10$\,mag, and $a=0.30$, $b=0.30$, and $c=0.35$. Since the main factor affecting the 
statistical error of the measurement is the number of stars 
within 1 mag above and below the TRGB, we adopted a ratio of star counts above/below the TRGB to a value typical for the LMC, that is 200/1000. We performed 10000 measurements 
on random generated luminosity functions, with different 
setups of the PN filter. We found that for a bin size value ($\mu$ in Eq. 4) from 0.2 to 0.4 mag and a $\sigma$ value from 0.01 to 0.04 
there is no significant difference for the distributions of the results. Figure 3 presents the results of the simulations 
for parameter $\mu$=0.2 and  $\mu$=0.4 mag. Those simulations convinced us that the PN filter can be properly used to measure 
the TRGB brightness in the LMC and SMC.

\subsection{TRGB measurement in the LMC and SMC} 
Utilising the method described in the previous subsection, we performed measurements in 19 fields in the LMC and SMC 
in the I, J, H, and K bands. Example of the measurement for field LMC 127 is presented in Figure 4. From the initially larger number 
of fields, we decided to use only measurements that were accurate and precise. 
The simplicity of the PN filter response makes it easy to reject measurements that provide some doubts. 
We decided to reject measurements if in the proximity of the anticipated edge there is no visual significant peak in the response 
of the PN filter, or if the peak 
has additional features, like a double maximum. We also rejected measurements if the magnitude of the maximum value changes 
significantly with changed $\mu$ parameter of the PN filter. It is worth noting that in all cases of measurement rejection, 
the Sobel filter response neither provided the possibility to measure the TRGB brightness. The statistical uncertainty of detection was 
estimated with a Bootstrap resampling method, and was smaller than 0.04 mag in all fields. 

Table 1 presents the measured values of the TRGB in the I, V, J and K bands for the selected fields. 
The TRGB measurements with the Sobel filter are not given in this paper, however they have been reported for the most of the fields earlier by G{\'o}rski et al. (2016).

\subsection{Color measurement}
To measure the color of the previously selected stars on the red giant branch, we selected stars of magnitude between measured 
brightness of the TRGB in the K band and 0.3 mag below the TRGB
(stars of magnitude $m$, where $m_\mathrm{K,\,TRGB} < m < m_\mathrm{K,\,TRGB} + 0.3$ mag). Next we applied the fitting function (Eq. 5) to the data, which 
consist of a Gaussian component representing RGB stars  and a second order polynomial approximating the stellar background. 
\begin{equation}
n(k)=a+b(k-k_0)+c(k-k_0)^2+\frac{N}{\sigma\sqrt{2\pi}}   \exp \left[ - \frac{(k-k_0)^2}{2 \sigma^2} \right]
\label{eq:RC}
\end{equation}
where $k$ is the color of the stars. In this paper we used colors (V-K), (V-H) and (J-K). Figure 5 presents an example of the fit 
applied to stars in field LMC127. Measured colors are reported in Table 2.

\section{Calibration of the TRGB}
In order to prepare the calibration of the absolute magnitude of the TRGB in terms of the unreddened colors of the 
red giant branch, we have to adopt a distance to the LMC and SMC, and interstellar extinction to both galaxies. We employ 
18.493 $\pm$ 0.047 mag distance modulus to the LMC (Pietrzy{\'n}ski et al. 2013), which is based on eight eclipsing binary systems 
and is the most precise and accurate (2\%) distance to this galaxy which has been determined so far.  
Fields analysed in this paper are located relatively close to the center of the LMC, and the geometrical depth of the LMC 
should not introduce any systematic error to our calibration.  Nevertheless it is one of the factors affecting the absolute magnitude 
of the TRGB in each field. To limit this effect, we applied the geometrical corrections calculated from the model of 
Van der Marel et al. (2002). Values of the geometrical corrections are reported in Table 2.

The adopted distance to the SMC is also based on the eclipsing binary method, but in this case only 5 systems were used 
to calculate the distance (Graczyk et al. 2014). From these 5 systems, we decided to take into account all systems except 
SMC113.3 4007 which is reported to be lying outside of the main body of the galaxy by a number of studies, and strongly 
affects the mean distance value (Matsunaga et al. 2011; Subramanian \& Subramaniam 2012). Based on those 4 systems, 
the adopted distance to the SMC is 19.003 $\pm$ 0.048 mag.

Relative reddenings for our fields were calculated based on the observed (V-I) colors of the red clump stars. The absolute zero point of the reddening was adopted from the Na I D1 line and atmospheric analysis of eclipsing binaries that are located in our fields (Graczyk et al. 2014; Graczyk et al. 2018). This approach is similar to the method used by Haschke, Grebel \& Duffau (2013), and will 
be discussed in the following section of this paper. The adopted E(B-V) reddening values are reported in Table 2. We use the
Schlegel et al. (1998) $R_V = 3.1$ reddening law and a ratio of total to selective absorption 
of A$_V$=1.08 R$_V$, A$_I$=0.568 R$_V$, A$_J$=0.288 R$_V$, A$_H$=0.178 R$_V$ and A$_K$=0.117 R$_V$.

To obtain the calibration of the TRGB brightness in terms of the color of the tip of the red giant branch, we used 
the least squares method to fit a first order polynomial to absolute magnitudes and unreddened colors calculated in the
previous sections. The fitted relations are of the form:
 \begin{equation}
 M_X = a (TC_z - TC_0)+b,
 \end{equation}
 where M$_X$ is the absolute magnitude of the TRGB in the band X, and TC$_z$ is the unreddened tip color, (V-K)$_0$, (V-H)$_0$ or (J-K)$_0$. 
 For the clarity of presented calibration formulas, we introduced TC$_0$ color shift (TC$_0$=3.8 for the (V-K)$_0$, TC$_0$=3.6 for the (V-H)$_0$ and TC$_0$=1.0 for the 
 (J-K)$_0$). Calculated $a$ and $b$ coefficients for I, J, H, and K bands for (V-K)$_0$, (V-H)$_0$ and (J-K)$_0$ tip colors are reported in Table 3. 
The fits are presented in Figures 6, 7 and 8.  
Here we explicitly present calibrations of the TRGB absolute magnitude in terms of the (V-K)$_0$ color of the tip. 

$$M_I =  0.09 \cdot ((V-K)_0-3.8) - 4.11 $$
$$M_J =  -0.28 \cdot ((V-K)_0-3.8) - 5.26 $$
$$M_H =  -0.37 \cdot ((V-K)_0-3.8) - 6.10 $$
$$M_K =  -0.48 \cdot ((V-K)_0-3.8) - 6.30 $$

\section{Discussion}

In this paper we present the calibration of the optical and near-infrared brightness of the TRGB in terms of the 
(V-K) and (V-H) colors for the first time. As a complementary calibration we provide a relation for the (J-K) color, which can be 
compared to the results obtained by  Serenelli et al. (2017) and  Hoyt et al. (2018). Calibrations based on the colors 
of the red giants should reduce potential systematic errors observed in calibrations based on the metallicity of the stars 
(Salaris \& Girardi 2005; G{\'o}rski et al. 2016; Serenelli et al. 2017). We note that in the color range of the presented 
calibrations, the absolute magnitude changes least in the optical I band (0.035 mag), and in the near-infrared K band the change 
of the brightness of the TRGB is the strongest (0.364 mag). This basic property is in agreement with all previously published 
calibrations. 

Our calibrations are valid only for the selected range of colors 3.4$<$(V-K)$<$4.1, 3.2$<$(V-H)$<$3.9, 0.94$<$(J-K)$<$1.07. 
We expect that extrapolating these calibrations to a wider color range will require the use of second order polynomials 
instead of the linear 
regression fits applied in this paper (Da Costa \& Armandroff 1990; Bellazzini \& Ferraro 2001; Serenelli et al. 2017). 
While formally we are able to perform higher order fits, the coefficients of the second order term in all fits are 
indistinguishable from zero within a 1$\sigma$ uncertainty. 

To measure the TRGB brightness we used the PN filter. For almost all analysed fields we were able to measure TRGB brightness 
with the Sobel filter as well. Using this measurement instead of the PN filter measurement doesn't change substantially any of our 
calibrations, but increases the spread. As an example, using Sobel filter measurements to calibrate the TRGB K band magnitude 
in terms of the (V-K) color yields coefficient values of a=-0.50 $\pm$ 0.04, b=-6.28 $\pm$ 0.01 with a spread $\sigma$=0.041 mag. 

Color measurement performed with fitting equation 5 secures precision and accuracy, since it distinguishes the main body 
of the red giant branch from the stellar background on the color-magnitude diagram. If we use the mean value of the color 
of the stars, the spread is significantly higher. We have to note that our convention of selecting stars to measure 
the color can introduce a significant offset because the color is effectively measured 0.15 mag below the TRGB. In fact 
it is the main cause of the discrepancy between our calibration, and the calibrations of Serenelli et al. (2017) 
and  Hoyt et al. (2018) visible in Figure 8. If instead of using measured color of the stars, we will simply take the difference of the 
measured TRGB brightness in the J band and measured TRGB brightness K band ($M_\mathrm{J,\,TRGB} - M_\mathrm{K,\,TRGB}$), we obtain relation virtually the same as Serenelli et al. (2017), 
but again with higher spread, $\sigma$=0.045 mag (Figure 9). 

The most important impact on our calibrations comes from the adopted distances to the LMC and SMC, and the adopted reddening. 
While the uncertainty on the LMC distance 
 is small compared to the other contributing uncertainties, the differential distance between LMC and SMC 
has a significant effect on our calibration. A 0.05 mag change of the adopted SMC distance modulus yields calibration $a$ and $b$ 
coefficient changes at the level of 2$\sigma$.  A corresponding effect can be attributed also to the reddening. 
Our reddening estimates for the analysed fields can be compared with 
Haschke, Grebel \& Duffau (2013) reddening maps, and with the values obtained from the reddening-law fitting to individual Cepheids 
in the LMC (Inno et al. 2016). Our mean reddening value agrees within 0.01 mag with values reported by Inno et al. (2016) 
with a standard deviation 0.04 mag. The reddening maps of Haschke, Grebel \& Duffau (2013) were obtained in 
a similar way that was used in this paper. In their case however, the color excess was calculated as the difference 
between the observed red clump color and the adopted theoretical value. If a zero point correction of 0.065 and 0.035 
is applied to the Haschke, Grebel \& Duffau (2013) reddening values to the LMC and SMC respectively, the values of reddening 
in each of our fields agree within 0.01 mag. 

\section{Summary and Conclusions}
Based on the measurements of the TRGB brightness in the optical I, and near-infrared J, H, and K bands in 19 separate fields 
in the Large and Small Magellanic Clouds we derived the calibrations of the TRGB absolute magnitude in terms of the (V-K)$_0$, 
(V-H)$_0$ and (J-K)$_0$ color of the red giant branch. All calibrations are expressed in the Landolt photometric system for the optical bands, 
in the 2MASS photometric system for near-infrared bands. The TRGB brightness measurements were performed with a simple edge 
detection technique which improves the accuracy and precision of the measurements. A reddening and geometrical correction 
was applied to each field separately, and the best distances available to both galaxies were adopted. 
The (V-K) color of the tip of the red giant branch is a robust tool allowing to calculate the absolute magnitude of TRGB 
with great precision. 

\acknowledgments
We would like to thank the anonymous referee for constructive and helpful comments.

The research leading to these results has received funding from the European Research Council (ERC) under the
European Union’s Horizon 2020 research and innovation program (grant agreement No 695099). WG, MG and DG
gratefully acknowledge financial support for this work from the Millenium Institute of Astrophysics (MAS) of the
Iniciativa Cientifica Milenio del Ministerio de Economia, Fomento y Turismo de Chile, project IC120009. We (WG,
GP and DG) also very gratefully acknowledge financial support for this work from the BASAL Centro de Astrofisica
y Tecnologias Afines (CATA) AFB-170002. We also acknowledge support from the IdP II 2015 0002 64 grant 
of the Polish Ministry of Science and Higher Education. M.G. gratefully acknowledges support from 
FONDECYT POSTDOCTORADO grant 3130361. Last not least, we are grateful to the OGLE and IRSF team members for providing data
of outstanding quality which made this investigation possible.

%-----------------------------------------
% LMC127 CMD
\begin{figure}[htb!]
\epsscale{0.7}
\plotone{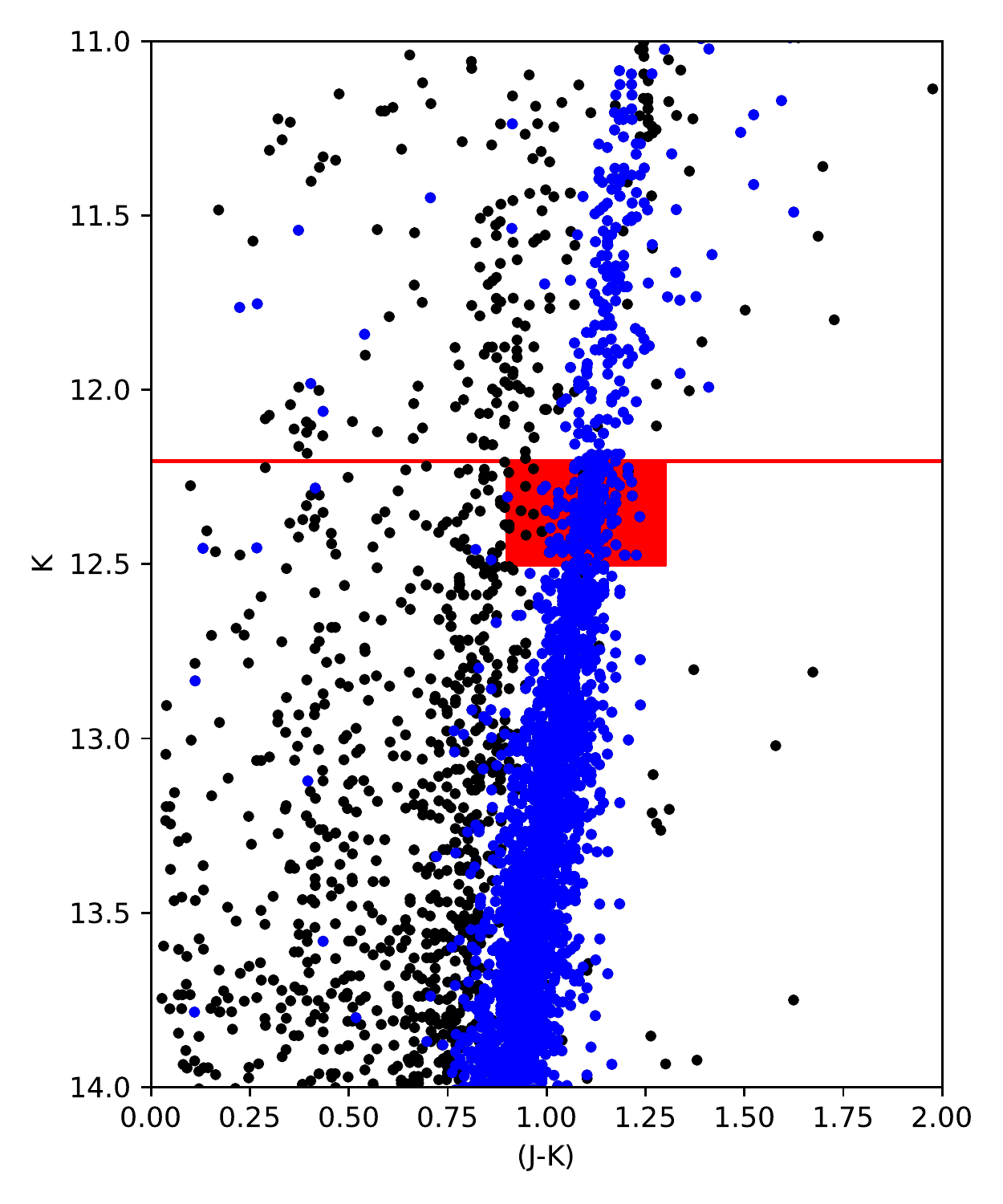}\\
\caption{Example of the color-magnitude diagram for field LMC127. Blue dots represent stars of the red giant branch, which were selected based on the (V-K) color and corresponding K band brightness. Red horizontal line marks the TRGB brightness measured with the PN filter, $m_{K,TRGB}$=12.204 mag. Red field square under 
the TRGB marks stars used to measure the color of the RGB.}
\label{fig:cmd}
\end{figure}

%-----------------------------------------
% Artificial LF
\begin{figure}[htb!]
\epsscale{0.5}
%\resizebox{0.5\linewidth}{!}{\plotone{Pic/artificial_a.png}}
%\resizebox{0.5\linewidth}{!}{\plotone{Pic/artificial_b.png}} \\
\resizebox{0.5\linewidth}{!}{\plotone{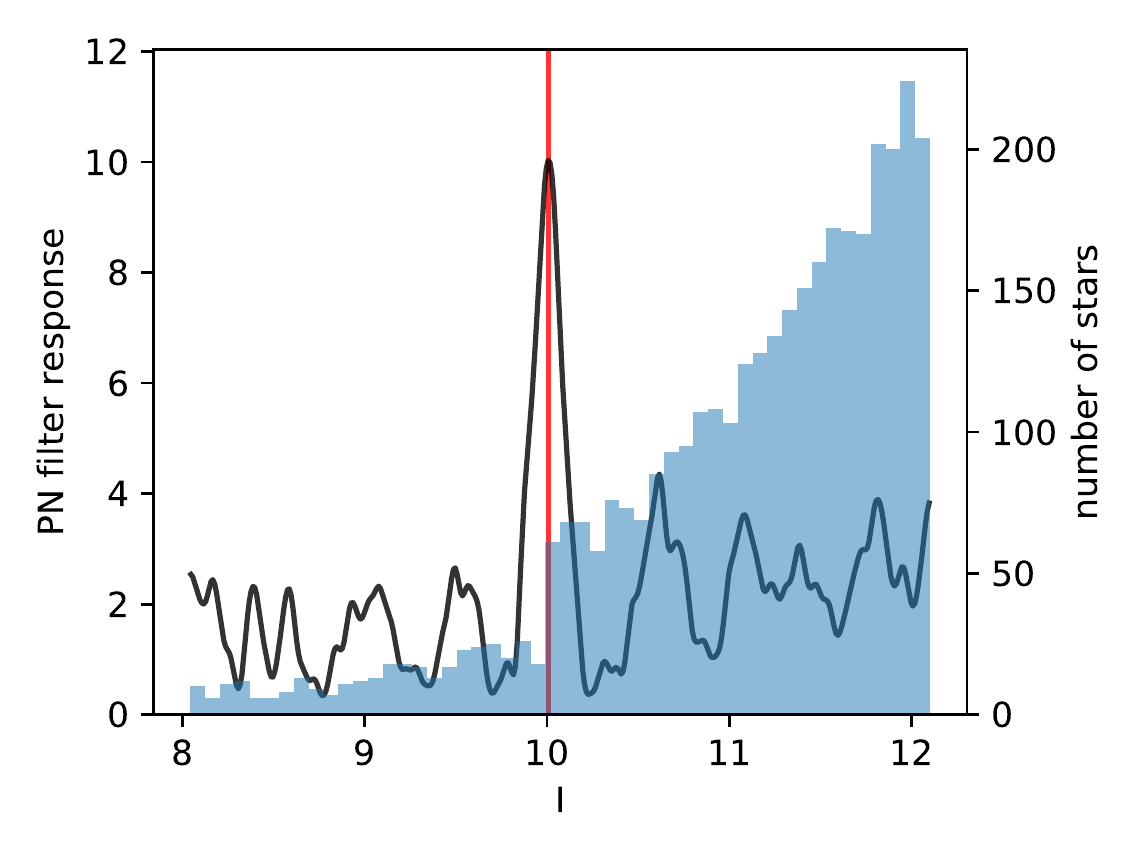}}
\resizebox{0.5\linewidth}{!}{\plotone{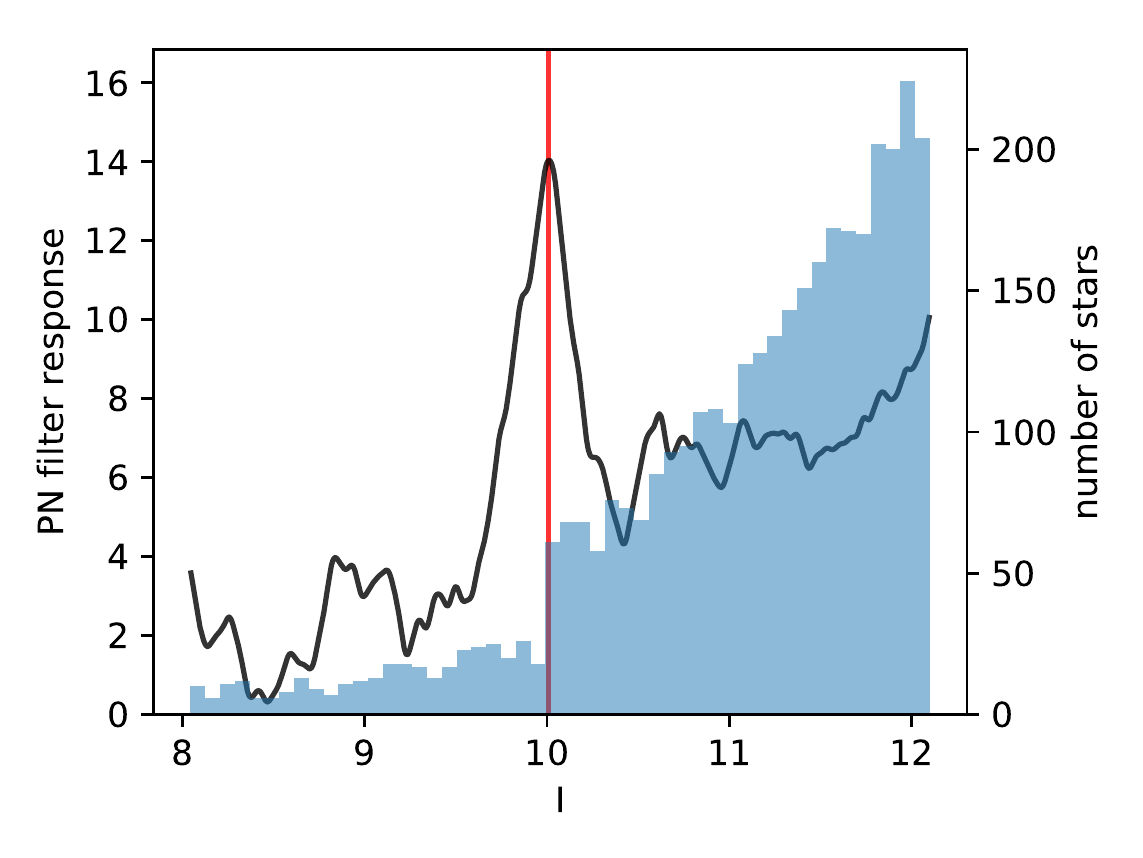}}

%\plotone{Pic/artificial_c.png}

\caption{Example of PN detection for an artificial luminosity function.  Both panels present power law distributions with a cut-off 
representing the TRGB at 10 mag. Power laws were generated according to Eq. 3 with slope parameters similar to those observed 
in the LMC (a=0.30, b=0.30, c=0.35). In this case the number of stars above/below the TRGB is 200 / 1000 which corresponds 
to typical values in our LMC fields. Left panel presents PN response with bin size 0.2 mag ($\mu$ parameter for Eq. 4). 
Right panel presents PN response with bin size 0.4 mag. It is clearly visible that for a power law distribution, with 
increasing bin size the PN response value has intrinsic rising trend 
with increasing magnitude, which can lead to a systematic measurement error, or even prevent 
the detection of the TRGB.}
\label{fig:art_lf}
\end{figure}
%-----------------------------------------
% Simulations

\begin{figure}[htb!]
\epsscale{0.5}
\resizebox{0.5\linewidth}{!}{\plotone{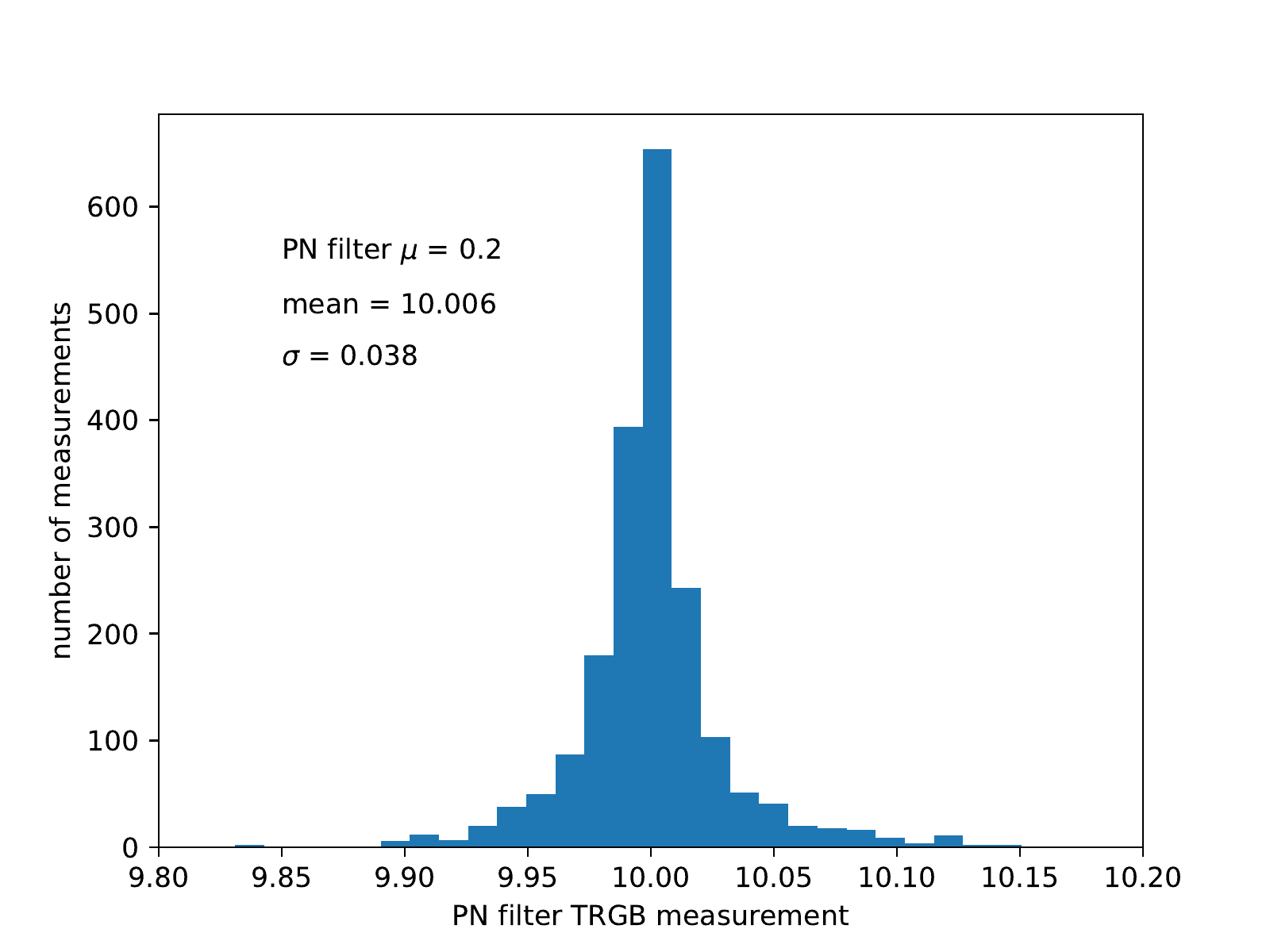}}
\resizebox{0.5\linewidth}{!}{\plotone{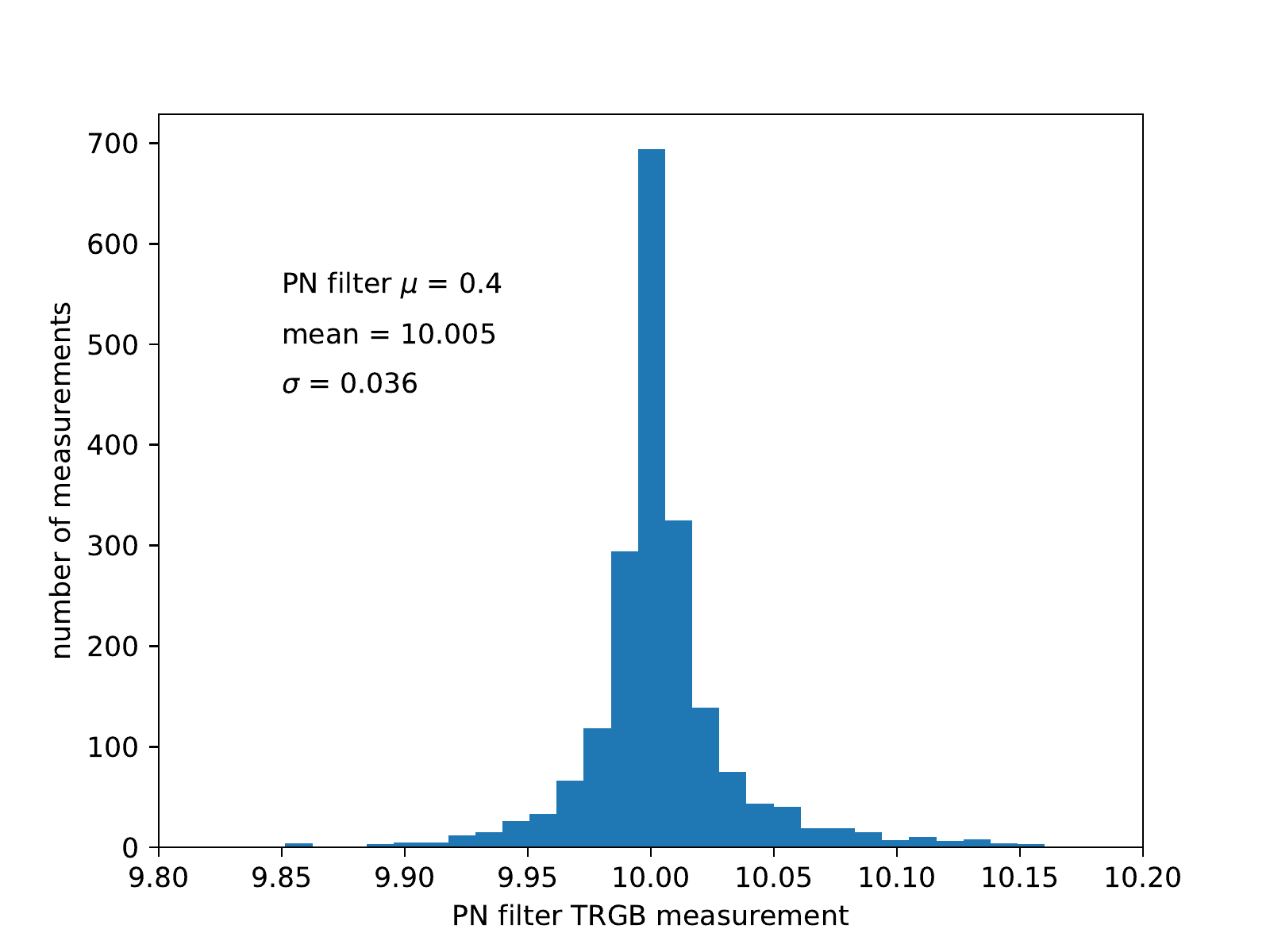}} \\

%\plotone{Pic/err_P01.png}
%\plotone{Pic/err_P05.png}
%\plotone{Pic/err_S.png}

\caption{
Distribution of the TRGB magnitude measurements of the PN filter with parameter $\mu$=0.2 mag (left panel) and $\mu$=0.4 mag 
(right panel). Each distribution was created by applying the PN filter to measure the TRGB magnitude for 2000 randomly generated 
artificial luminosity functions according to Eq. 3. Slope parameters were set to a=0.30, b=0.30, c=0.35, and number of stars 
above/below the TRGB were set to 200 / 1000. These values are typical for our analysed fields in the LMC, and were found with 
the MLA technique. The TRGB magnitude (m$_{\mathrm{TRGB}}$) was set to 10 mag. The presented distributions prove that there is no systematic 
error related to application of the PN filter within used $\mu$ parameter. Both distributions have a similar standard deviation 
and no shift of the mean value of the distribution is observed.
}
\label{fig:art_lf}
\end{figure}
%-----------------------------------------

%-----------------------------------------
% OK filed

\begin{figure}[htb!]
\epsscale{0.5}
\resizebox{0.5\linewidth}{!}{\plotone{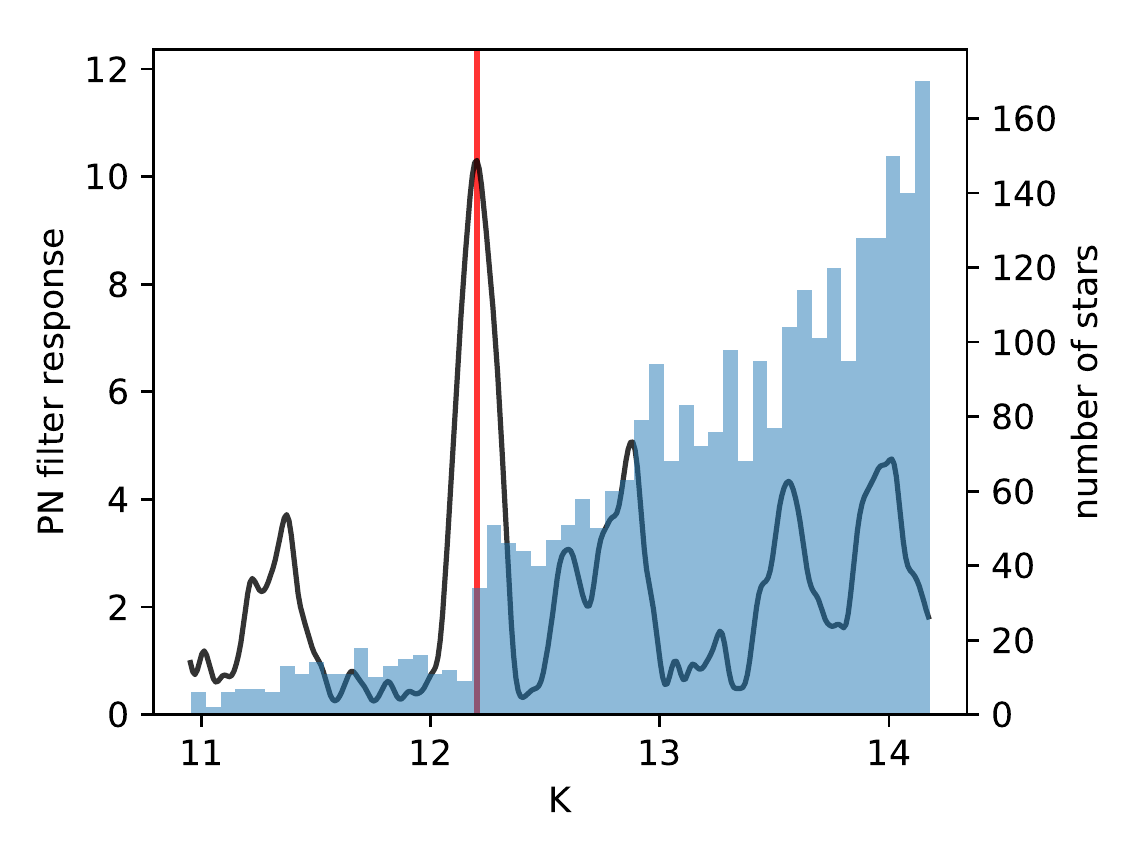}}
\resizebox{0.5\linewidth}{!}{\plotone{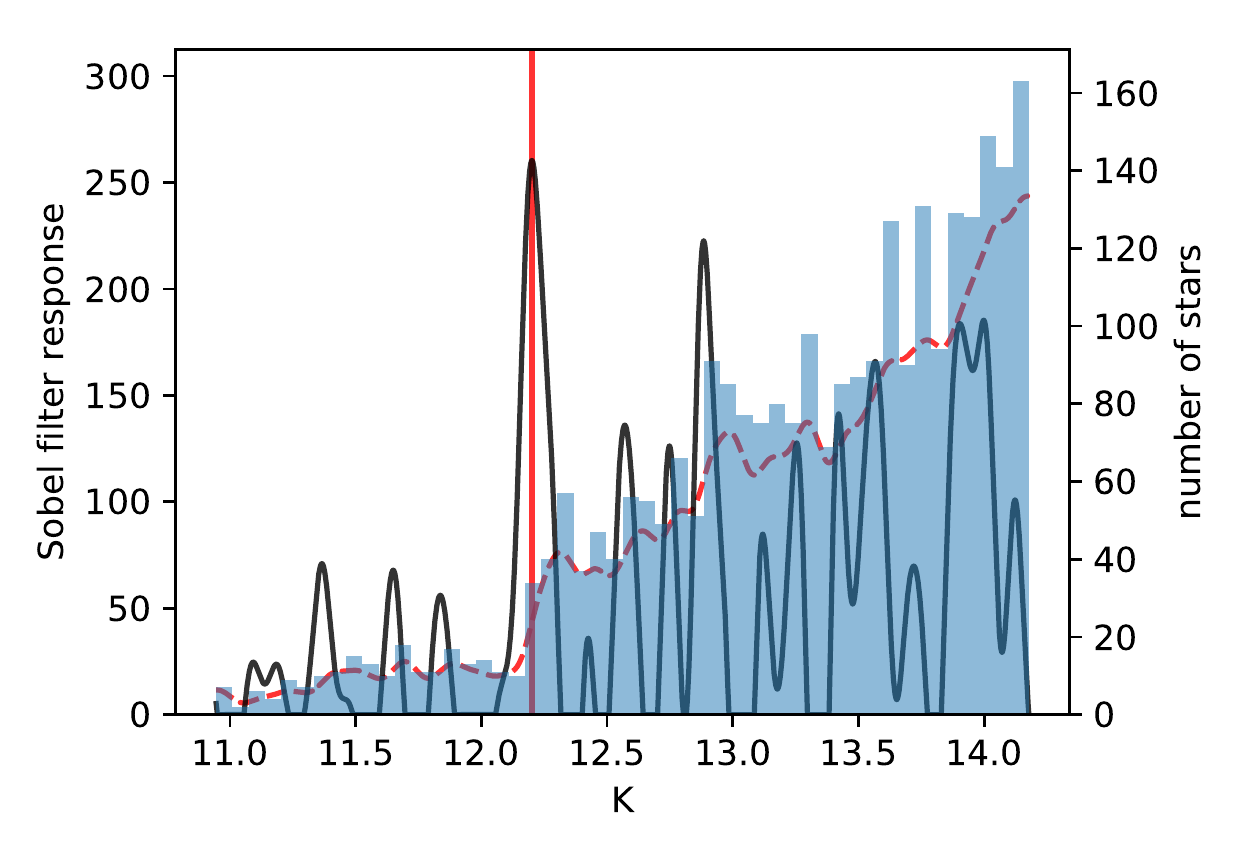}}

%\plotone{Pic/artificial_c.png}

\caption{Example of the PN filter (left panel) and Sobel filter (right panel) application for the field LMC127. Both filters 
provide results which are consistent within 0.01 mag. Red vertical line marks the measured magnitude of the TRGB.}
\label{fig:okPN}
\end{figure}

%-----------------------------------------

\newpage

\begin{figure}[htb!]
\epsscale{0.6}
\plotone{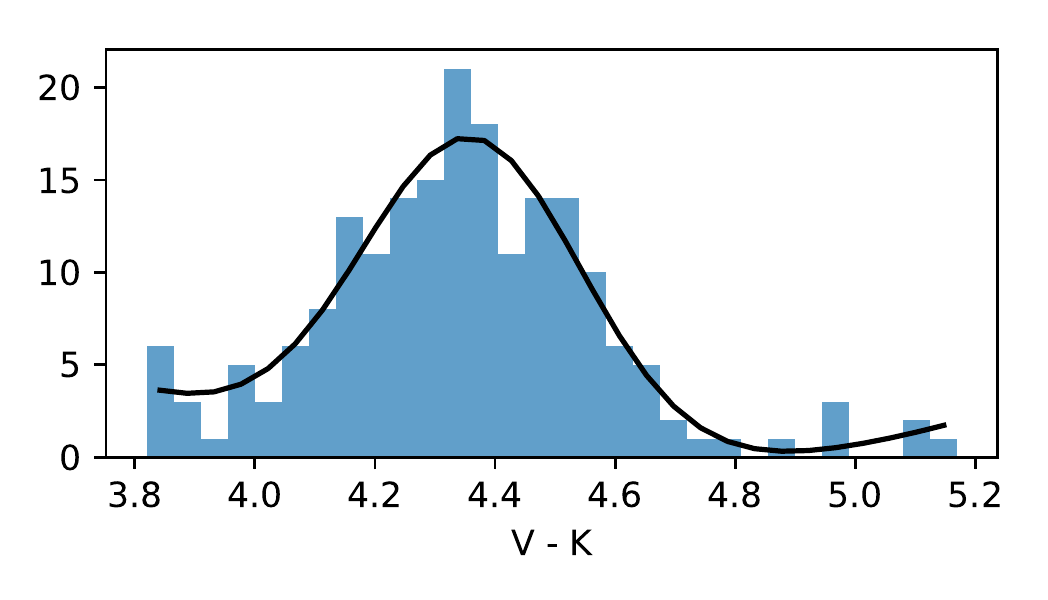}\\
\caption{Example of the fit to the color of the stars in the field LMC127 according to Eq. 5. Stars were selected from the 
red giant branch within the magnitude $m$, $m_\mathrm{J,\,TRGB} - m_\mathrm{K,\,TRGB}$+0.3 mag. }
\label{fig:colorfit}
\end{figure}
%-----------------------------------------

%-----------------------------------------
%V-K
\begin{figure}[htb!]
\epsscale{1}
\plotone{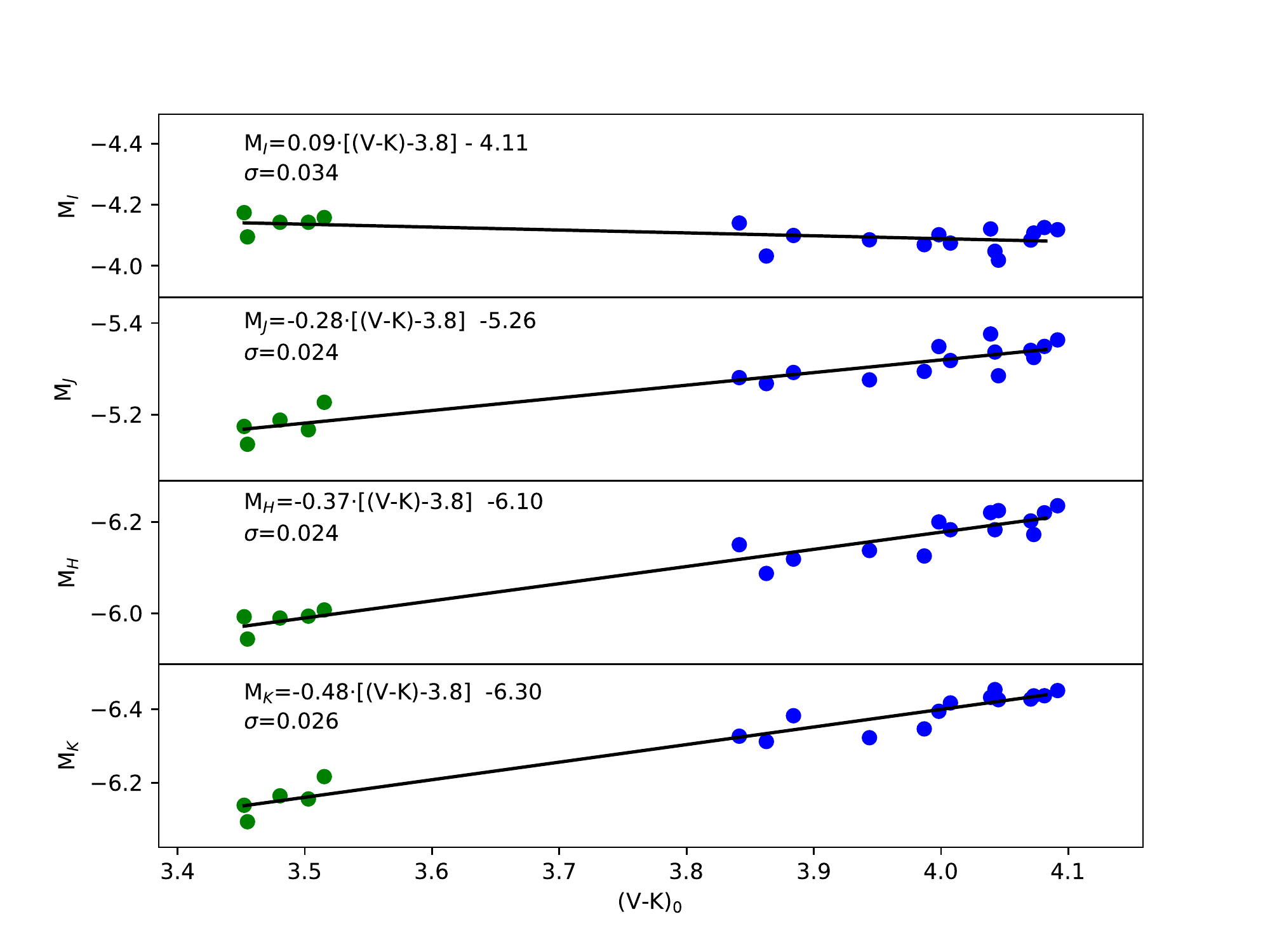}\\
\caption{Absolute magnitudes of the TRGB as a function of the tip (V-K)$_0$ color. Green points come from 5 fields in the SMC, blue 
points from 14 fields in the LMC. Black solid line is the best fit to the
data.}
\label{fig:vk}
\end{figure}
%-----------------------------------------
%V-H
\begin{figure}[htb!]
\epsscale{1}
\plotone{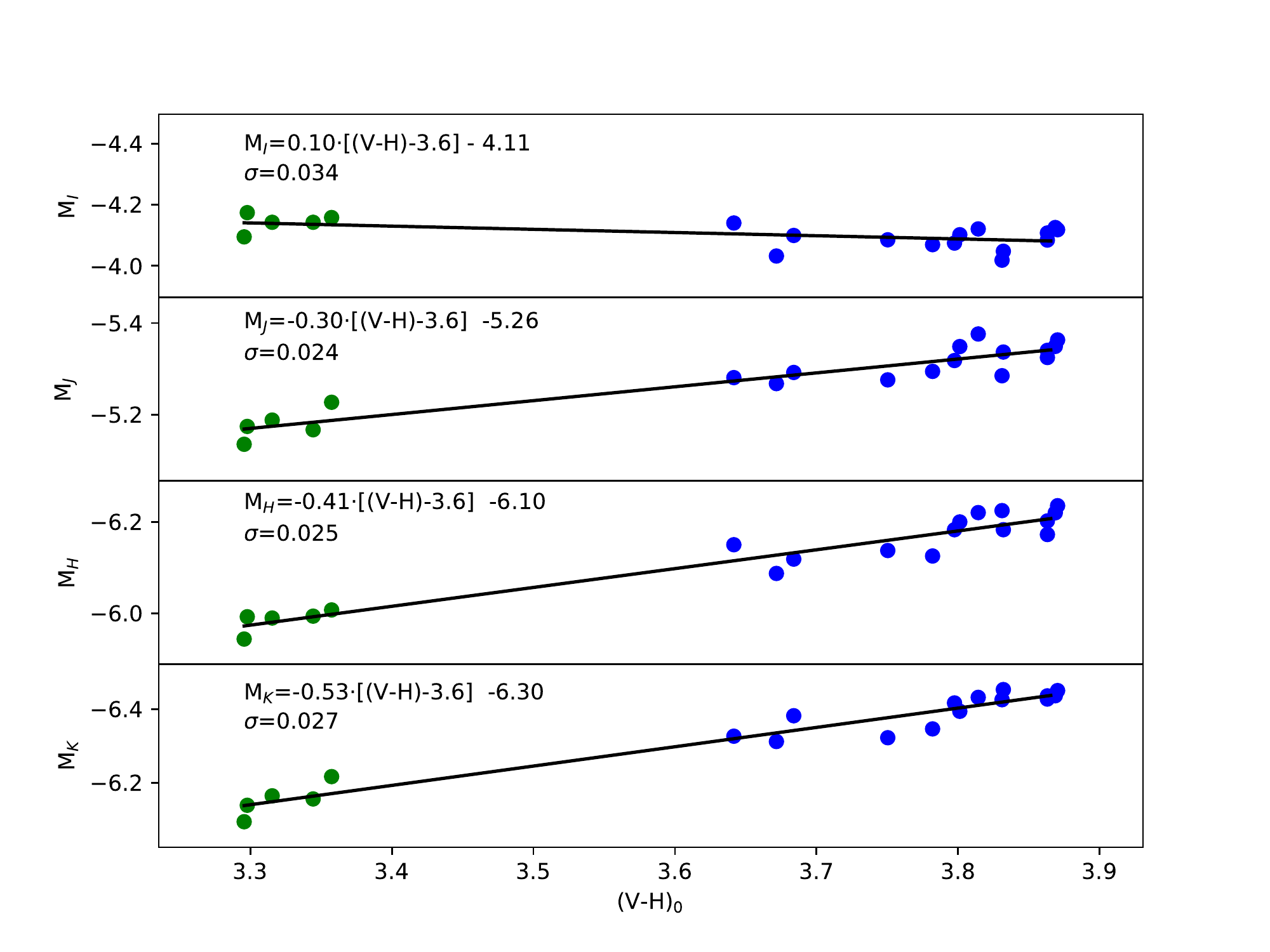}\\
\caption{Absolute magnitudes of the TRGB as a function of the tip (V-H)$_0$ color. Green points come from 5 fields in the SMC, 
blue points from 14 fields in the LMC. Black solid line is the best fit to the
data.}
\label{fig:vh}
\end{figure}
%-----------------------------------------
% J-K
\begin{figure}[htb!]
\epsscale{1}
\plotone{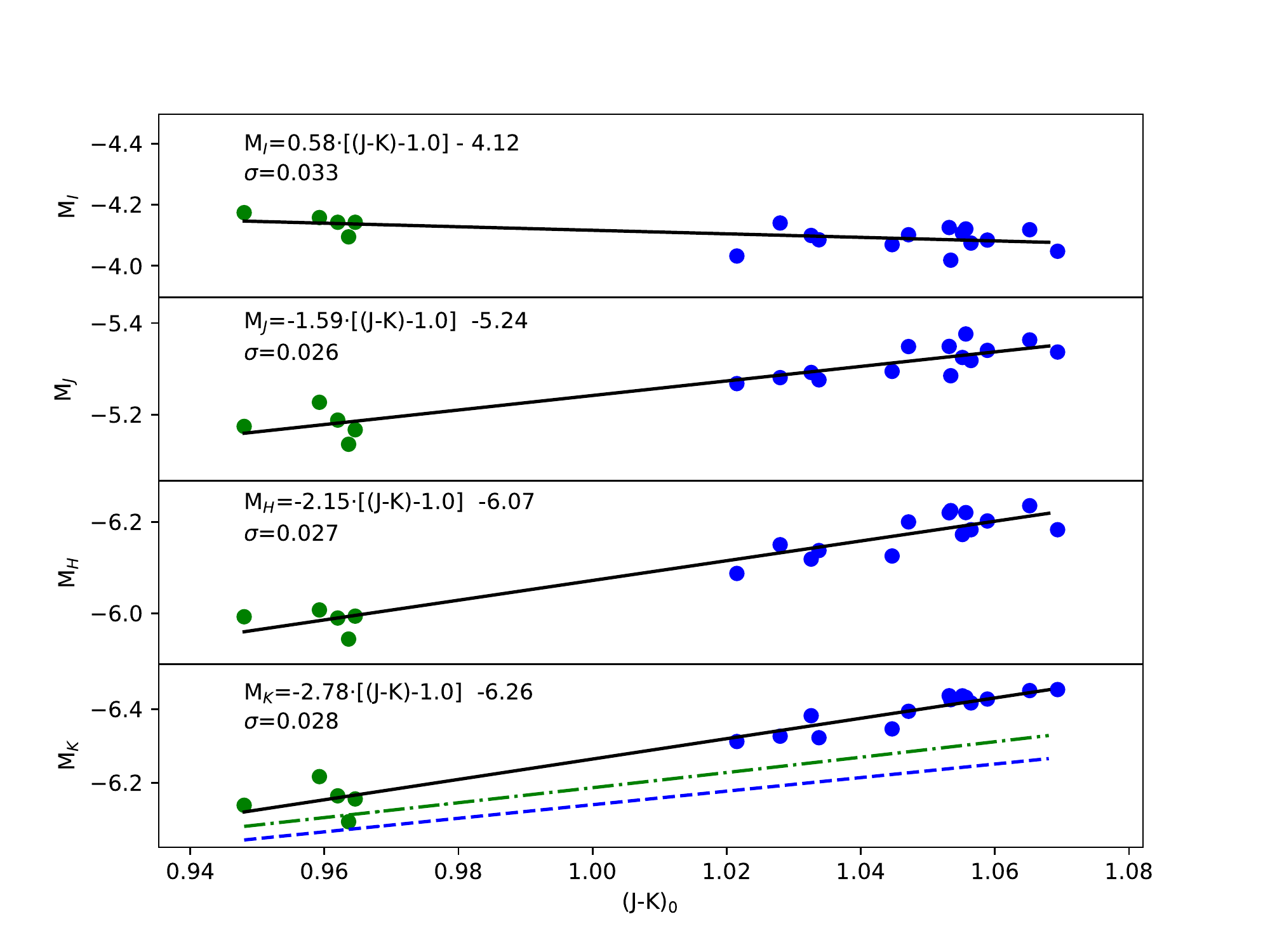}\\
\caption{Absolute magnitudes of the TRGB as a function of the tip (J-K)$_0$ color. Green points come from 5 fields in the SMC, blue 
points from 14 fields in the LMC. Black solid line is the best fit to the
data. The blue dashed line is the calibration of Hoyt et al. (2018). The green dashed-dot line is the calibration of 
Serenelli et al. (2018). }

\label{fig:jk}
\end{figure}
%-----------------------------------------

%-----------------------------------------
% J-K

\newpage

\begin{figure}[htb!]
\epsscale{1}
\plotone{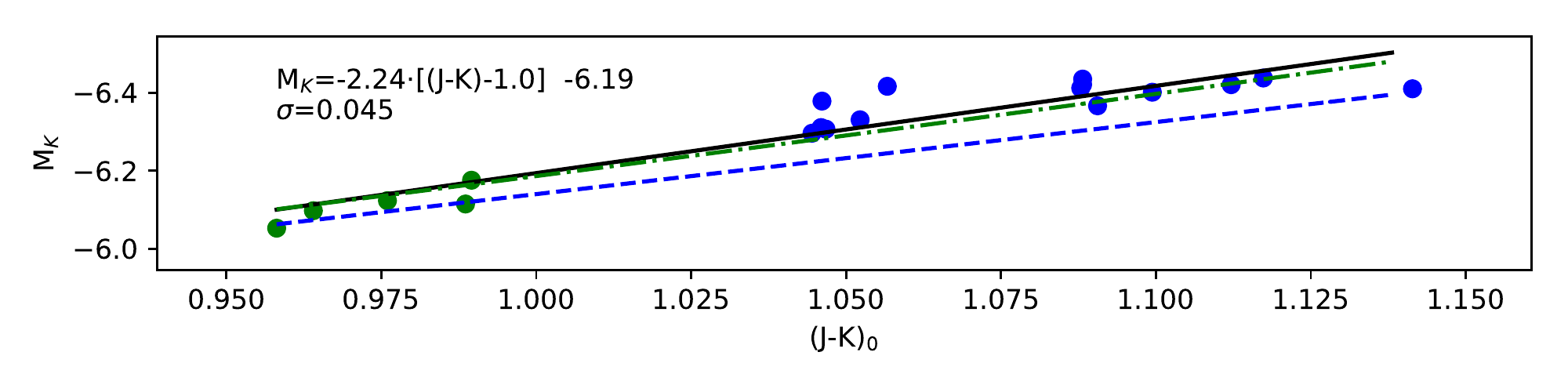}\\
\caption{Absolute magnitudes of the TRGB as a function of tip color calculated as the difference of the TRGB brightness in the J and K band. Green points are 5 fields in the SMC, and blue points are 14 fields in the LMC. Black solid line is the best fit to the
data. This approach yields different values of calibration coefficients, and makes the calibration consistent with calibration of Hoyt et al. (2018) - blue dashed line, 
and with Serenelli et al. (2018) - green dash-dot line.}
\label{fig:cal_v2}
\end{figure}
%-----------------------------------------

%----------- TABLES ---------------------
%---------------------------------------------

\begin{deluxetable}{cccccccc}
\tablewidth{0pc}
\tablecaption{Summary information on the 19 analysed fields in the LMC and SMC. For each field the coordinates of the center, and the TRGB brightness in I, J, H, and K bands are given.}

\tablehead{
\colhead{Field} & \colhead{RA} & \colhead{DEC} & \colhead{I$_{TRGB}$} & \colhead{J$_{TRGB}$} & \colhead{H$_{TRGB}$} & \colhead{K$_{TRGB}$}  \\}

\startdata
LMC100 & 5:19:02.2  & -69:15:07  & 14.581 & 13.254 & 12.343 & 12.103 \\
LMC102 & 5:19:05.7 & -68:03:46  & 14.661 & 13.382 & 12.459 & 12.252 \\
LMC103 & 5:19:02.9 & -69:50:26  & 14.615 & 13.255 & 12.353 & 12.104 \\
LMC111 & 5:12:36.0 & -69:14:50  & 14.714 & 13.328 & 12.342 & 12.114 \\
LMC112 & 5:12:21.5 & -69:50:21  & 14.602 & 13.273 & 12.382 & 12.093 \\
LMC116 & 5:07:03.6 & -67:28:25  & 14.682 & 13.350 & 12.493 & 12.247 \\
LMC120 & 5:05:39.8 & -69:50:28  & 14.643 & 13.302 & 12.426 & 12.179 \\
LMC126 & 5:00:02.4 & -68:39:31  & 14.620 & 13.313 & 12.442 & 12.153 \\
LMC127 & 4:59:33.6 & -69:14:54   & 14.638 & 13.325 & 12.416 & 12.204 \\
LMC161 & 5:25:32.5 & -69:14:59 & 14.624 & 13.267 & 12.373 & 12.154 \\
LMC162 &  5:25:43.3 & -69:50:24  & 14.579 & 13.234 & 12.323 & 12.085 \\
LMC163 &  5:25:52.2 & -70:25:50  & 14.648 & 13.298 & 12.392 & 12.134 \\
LMC169 & 5:32:22.8 & -69:50:26  & 14.691 & 13.284 & 12.392 & 12.095 \\
LMC170 & 5:32:48.1 & -70:25:53   & 14.600 & 13.242 & 12.358 & 12.123 \\
\hline
SMC101 & 0:50:03.5 & -72:33:03  & 15.017 & 13.894 & 13.062 & 12.871 \\
SMC108 & 0:57:31.5 & -72:09:29  & 14.972 & 13.901 & 13.055 & 12.894 \\
SMC105 & 0:57:50.2 & -72:44:35  & 15.081 & 13.955 & 13.113 & 12.945 \\
SMC106 & 0:58:06.7 & -73:20:21  & 14.995 & 13.904 & 13.051 & 12.875 \\
SMC113 & 1:05:02.8 & -72:09:32 & 14.994 & 13.851 & 13.042 & 12.817 \\
\enddata
\label{tab:smc}
\end{deluxetable}

%--------------------------------------------------------------------------------------------

\begin{deluxetable}{cccccccccc}
\tablewidth{0pc}
\tabletypesize{\small}
\tablecaption{TRGB absolute magnitude, unreddened color of the tip, geometric correction and reddening in 19 fields in the LMC and SMC.}

\tablehead{
\colhead{Filed} & \colhead{I$_{TRGB}$} & \colhead{J$_{TRGB}$} & \colhead{H$_{TRGB}$} & \colhead{K$_{TRGB}$} & \colhead{(V-K)$_0$} & \colhead{(V-H)$_0$} & \colhead{(J-K)$_0$} &
\colhead{geometric} & \colhead{E(B-V)} \\
\colhead{} & \colhead{} & \colhead{} & \colhead{} & \colhead{} & \colhead{} & \colhead{} & \colhead{}  & \colhead{correction} 

}
\startdata
LMC100 & -4.087 & -5.318 & -6.192 & -6.411 & 4.107 & 3.893  & 1.058 & -0.004 & 0.110 \\
LMC102 & -4.101 & -5.250 & -6.122 & -6.300 & 3.870 & 3.668 & 1.033 & -0.028 & 0.149 \\
LMC103 & -4.047 & -5.310 & -6.174 & -6.402 & 4.095 & 3.887  & 1.063 & 0.004 & 0.111 \\
LMC111 & -3.972 & -5.250 & -6.194 & -6.399 & 4.085 & 3.868  & 1.060 & 0.002 & 0.124 \\
LMC112 & -4.073 & -5.296 & -6.146 & -6.412 & 4.093 & 3.882  & 1.059 & 0.009 & 0.122 \\
LMC116 & -4.000 & -5.240 & -6.062 & -6.288 & 3.878 & 3.686  & 1.024 & -0.026 & 0.106 \\
LMC120 & -4.040 & -5.268 & -6.101 & -6.323 & 3.997 & 3.792  & 1.047 & 0.015 & 0.130 \\
LMC126 & -4.069 & -5.266 & -6.094 & -6.358 & 3.896 & 3.695  & 1.035 & 0.005 & 0.127 \\
LMC127 & -4.057 & -5.250 & -6.113 & -6.299 & 3.953 & 3.759  & 1.035 & 0.017 & 0.137 \\
LMC161 & -4.091 & -5.332 & -6.181 & -6.374 & 3.977 & 3.781  & 1.043 & -0.010 & 0.134 \\
LMC162 & -4.078 & -5.331 & -6.207 & -6.425 & 4.122 & 3.899  & 1.071 & -0.002 & 0.105 \\
LMC163 & -4.041 & -5.290 & -6.157 & -6.393 & 4.025 & 3.814 & 1.060 & -0.014 & 0.116 \\
LMC169 & -4.008 & -5.305 & -6.155 & -6.428 & 4.070 & 3.858 & 1.074 & -0.007 & 0.126 \\
LMC170 & -4.093 & -5.351 & -6.196 & -6.409 & 4.047 & 3.822 & 1.057 & -0.020 & 0.115 \\
\hline
SMC101 & -4.085 & -5.148 & -5.956 & -6.134 & 3.539 & 3.370 & 0.972 & -- & 0.069 \\
SMC108 & -4.119 & -5.135 & -5.960 & -6.109 & 3.507 & 3.349 & 0.958 & -- & 0.063 \\
SMC105 & -4.035 & -5.094 & -5.910 & -6.063 & 3.518 & 3.354 & 0.975 & -- & 0.077 \\
SMC106 & -4.087 & -5.128 & -5.961 & -6.126 & 3.558 & 3.396 & 0.974 & -- & 0.058 \\
SMC113 & -4.095 & -5.184 & -5.972 & -6.186 & 3.583 & 3.420 & 0.971 & -- & 0.062 \\
\enddata
\end{deluxetable}

%--------------------------------------------------------------------------------------------
%--------------------------------------------------------------------------------------------

\begin{deluxetable}{cccc}
\tablewidth{0pc}
\tablecaption{Equation 5 calibration formula coefficients}

\tablehead{ \colhead{Band } &  \colhead{(V-K)$_0$ -3.8} &  \colhead{(V-H)$_0$ -3.6} &  \colhead{(J-K)$_0$ -1.0}   }

\startdata
%Band &  [(V-K)-3.8] & [(V-H)-3.6] & [(V-I)-1.7] & [(J-K) -1.0]   \\
%\hline
 %\hline

 \vspace{-0.8cm} M$_I$ \\        
                                          & a=0.094 $\pm$ 0.034 & a=0.104 $\pm$ 0.037   & a= 0.579  $\pm$ 0.194 \\
                                          & b=-4.107 $\pm$ 0.008 & b=-4.109  $\pm$ 0.008 & b= -4.116  $\pm$ 0.009 \\                                
\hline \\
\vspace{-0.8cm}   M$_J$     \\
                                   & a=-0.275 $\pm$ 0.023 & a=-0.302  $\pm$  0.027 & a= -1.586 $\pm$ 0.152 \\
                                   & b=-5.264 $\pm$ 0.006 & b=-5.261  $\pm$ 0.006 & b=-5.241  $\pm$ 0.007 \\

\hline \\
\vspace{-0.8cm}  M$_H$  &&&  \\
                                   & a=-0.374 $\pm$ 0.024 & a=-0.411 $\pm$ 0.027  &  a= -2.154 $\pm$ 0.162 \\
                                  & b=-6.103 $\pm$ 0.006 & b=-6.097 $\pm$ 0.006  &  b= -6.072 $\pm$ 0.008  \\

\hline \\
\vspace{-0.8cm} M$_K$  &&&  \\
                                 & a=-0.479 $\pm$ 0.026& a=-0.527 $\pm$ 0.030  &   a=-2.779   $\pm$ 0.166  \\
                                & b=-6.304 $\pm$ 0.006 & b=-6.298 $\pm$  0.007 & b=-6.264  $\pm$0.008  \\

\enddata
\label{tab:abc}
\end{deluxetable}

\end{document}